\documentclass[aps,prb,twocolumn,showpacs,amsfonts,amsmath,amssymb,groupaddress,tightlines]{revtex4-1}%

\usepackage{graphicx}

\begin{document}


\newcommand{\FeNi}{Ba(Fe$_{0.954}$Ni$_{0.046}$)$_2$As$_2$ }
\newcommand{\FeCo}{Ba(Fe$_{0.926}$Co$_{0.074}$)$_2$As$_2$ }

\title{Magneto-optical study of Ba(Fe$_{1-x}$T$_{x}$)$_2$As$_2$ (T=Co, Ni) single crystals \\ irradiated with heavy-ions}

\author{R.~Prozorov}
\email[Corresponding author: ]{prozorov@ameslab.gov}

\author{M.~A.~Tanatar}

\author{B.~Roy}

\author{N.~Ni}

\author{S.~L.~Bud'ko}

\author{P.~C.~Canfield}
\affiliation{Ames Laboratory and Department of Physics \& Astronomy, Iowa State University, Ames, IA 50011}

\author{J.~Hua}

\author{U.~Welp}

\author{W.~K.~Kwok}
\affiliation{Materials Science Division, Argonne National Laboratory, Argonne, Illinois 60439}

\date{19 January, 2010}

\begin{abstract}
Optimally doped single crystals of Ba(Fe$_{1-x}$T$_x$)$_2$As$_2$ (T=Co, Ni) were irradiated with 1.4 GeV $^{208}$Pb$^{56+}$ ions at fluences corresponding to matching fields of $B_{\phi}=0.1,\,0.5,\,1$ and $2$ T. Magneto-optical imaging has been used to map the distribution of the magnetic induction in the irradiated samples. The imaging is complemented by the magnetization measurements. The results show a substantial enhancement of the apparent critical current densities as revealed by the much larger Bean penetration fields and an increase of the hysteretic magnetization. However, the effect depends on the compound, temperature and applied magnetic field. In \FeCo crystals, at 15 K and low fields, the enhancement appears to scale with the irradiation dose at a rate of about 0.27\,MA$\cdot$cm$^{-2}$T$^{-1}$, whereas in \FeNi crystals, higher irradiation doses are less effective. Our results suggest that moderate irradiation with heavy ions is a an effective way to \emph{homogeneously} enhance the current-currying capabilities of pnictide superconductors.
\end{abstract}

\pacs{74.70.Xa, 74.25.Ha, 74.25.Qt, 74.25.Sv}


\maketitle

Studies of irreversible vortex properties and critical currents are important for any superconductor due to potential applications as well as fundamental questions related to anisotropy and pairing mechanisms.\cite{Brandt1995,Blatter1994} In the case of iron-based high-$T_c$ superconductors, there is an additional interest in comparing their behavior to the cuprates and MgB$_2$.\cite{Grant2008,Chu2009} Pinning of Abrikosov vortices depends both on the vortex structure and on the pinning potential. Whereas the pairing mechanism remains an open question, there is evidence that defects extended along the vortex direction cause a significant pinning enhancement. In particular, naturally occurring planar and linear defects in the form of structural/magnetic domains\cite{Tanatar2009} and their intersections in the underdoped regime of the Ba(Fe$_{1-x}$Co$_x$)$_2$As$_2$ (FeCo-122) system were shown to constrain the motion of vortices either by trapping them\cite{Prozorov2009b} or repelling due to an enhanced superfluid density at the boundaries.\cite{Kalisky2009} It is possible to introduce linear defects in the form of columnar tracks by heavy-ion irradiation, which was very successful in the case of high-$T_c$ cuprates.\cite{Brandt1995,Blatter1994}
Since most of the cuprates, with the possible exception of Y-Ba-Cu-O, are highly anisotropic, the effectiveness of pinning by extended defects is reduced due to the formation of "pancake" vortices. However, columnar defects still enhance interlayer Josephson coupling by suppressing the thermal fluctuations of the pancakes.\cite{Koshelev1996} In low-anisotropy, three-dimensional FeCo-122 crystals, a pronounced effect of heavy-ion irradiation has been convincingly demonstrated by Nakajima \emph{et al.}\cite{Nakajima2009} Neutron irradiation, producing point-like defects, resulted in a somewhat smaller enhancement of pinning.\cite{Zehetmayer2009a} An increase of $J_c$ due to columnar tracks was reported in polycrystalline NdFeAsO$_{0.85}$.\cite{Moore2009}

\begin{figure}[ht]
\begin{center}
\includegraphics[width=8.5cm]{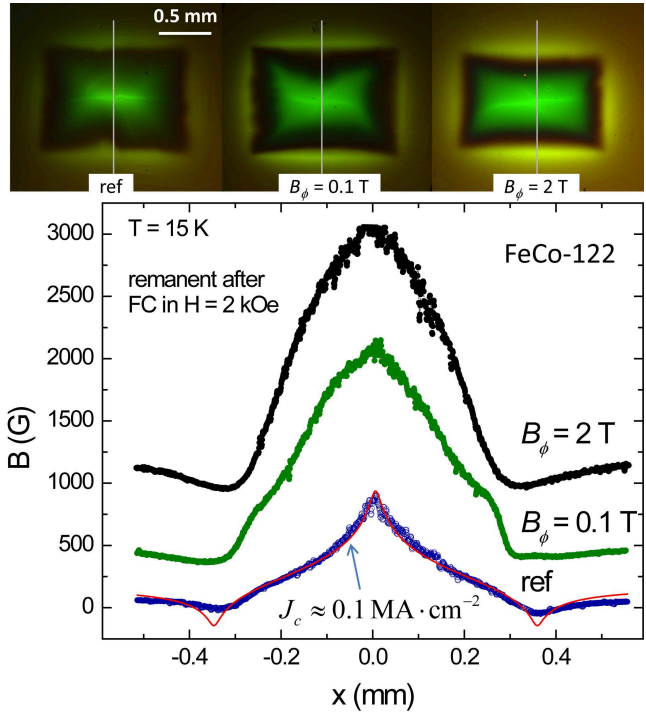}
\caption{(Color online) MO maps of the remanent (trapped) flux after cooling in 2 kOe to 15 K and turning the field off, obtained simultaneously in three FeCo-122 crystals: reference, $B_{\phi}=0.1$ T and $B_{\phi}=2$ T. Lower panel: $B(r)$ profiles (shifted vertically for clarity) measured along the directions shown by vertical lines in the upper panel. Solid line is a fit to Eq.\,(\ref{eq1}).}
\label{fig1}
\end{center}
\end{figure}

There are many questions with regard to the influence of heavy-ion irradiation on physical properties of a superconductor. The introduction of defects leads to the enhancement of pinning, but it may also suppress superconductivity, especially in unconventional superconductors and there is mounting evidence that pnictides are such. Other questions pertain to the mesoscopic (in)homogeneity of the irradiated samples, the effectiveness of different irradiation doses and determining the field and temperature range where irradiation is most effective. Here we present results of magneto-optical studies of \FeCo and \FeNi single crystals irradiated with 1.4 GeV $^{208}$Pb$^{56+}$  ions. We observe a \emph{significant}, \emph{uniform enhancement} of vortex pinning at moderate irradiation doses. Higher doses are still effective in FeCo-122, but not in FeNi-122.

Single crystals of Ba(Fe$1_x$T$_x$)$_2$As$_2$ (T=Co ($x=0.074$), Ni ($x=0.046$)) were grown out of FeAs flux using a high temperature solution growth technique. Details of the growth and physical characterization can be found elsewhere.\cite{Ni2008b,Canfield2009,Bud'ko2009} X-ray diffraction, resistivity, magnetization, magneto-optics and wavelength dispersive spectroscopy (WDS) elemental analysis have all shown good quality single crystals at these optimal dopings with a small variation of the dopant concentration over the sample and sharp superconducting transitions, $T_c = 23$ K for FeCo-122 and $T_c = 18$ K for FeNi-122. Vortex properties of unirradiated FeCo-122 crystals from the same batch were described previously.\cite{Prozorov2008,Prozorov2009c} To examine the effects of irradiation, $\sim 2 \times 0.5 \times 0.02 - 0.05$ mm$^3$ single crystals were selected and then cut into three or four pieces preserving the width and the thickness. Hence, the results reported here compare sets of samples, where the samples in each set are parts of \emph{the same original, large crystal}. Several such sets were prepared and for each a reference (ref) piece was kept unirradiated. The thickness was chosen in the range of  $20 - 50 \mu$m to be smaller than the penetration depth of the ions, $~\sim 60-70 \mu$m. Irradiation with 1.4 GeV $^{208}$Pb$^{56+}$ ions was performed at the Argonne Tandem Linear Accelerator System (ATLAS) with ion flux of $\sim 5\times 10^{11}$ ions$\cdot$s$^{-1}\cdot$m$^{-2}$. In each run, the actual total dose was recorded. The density of defects, $n$, created by the irradiation is usually expressed in terms of the matching field, $B_{\phi}=\Phi_0n$, which is obtained assuming one columnar track per ion, each occupied by an Abrikosov vortex with flux quanta, $\Phi_0 \approx 2.07 \times 10^{-7}$ G$\cdot$cm$^2$, so that the mean distance between the columnar tracks is $a\approx\sqrt{\Phi_0/B_{\phi}}$. Here we studied samples with $B_{\phi}=0.1,\,0.5,\,1$ and $2$\,T, which correspond to $a\approx 1438,\,643,\,455$ and $322$ \AA, respectively.

Magneto-optical (MO) imaging was performed in a $^{4}$He optical flow-type cryostat utilizing the Faraday rotation of linearly polarized light in a bismuth - doped iron-garnet ferrimagnetic indicator film with in-plane magnetization.\cite{Vlasko-Vlasov1999,Jooss2002a}. The spatial resolution of the technique is about 3 $\mu$m. In all images, the intensity is proportional to the local value of $B_z$ (perpendicular to the sample surface). Different colors (online) indicate different directions of the magnetic field. Crystals were placed on a polished copper disc and imaged simultaneously, thanks to their identical thicknesses, allowing for identical experimental conditions for each sample in a given set. The critical current densities were estimated from magnetization measurements (performed with a \emph{Quantum Design} MPMS) using the Bean model\cite{Bean1962}, in which for a rectangular slab with the dimensions of $2d <2a \leq 2b$, $J_c\mathrm{ [A/cm^2]}=20M\mathrm{ [emu]}/\left[aV(1-a/(3b))\right]$, where all dimensions are in cm, $M$[emu] is the magnetic moment and $V=8abd$ is the sample volume. Similar values were obtained by fitting the flux profiles.

\begin{figure}[tb]
\begin{center}
\includegraphics[width=8.6cm]{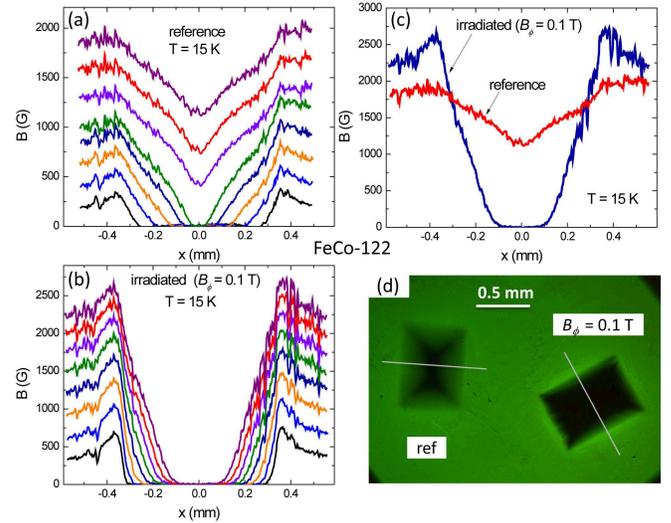}
\caption{(Color online) Flux penetration at $T=15$\,K in reference (a) and irradiated with $B_{\phi}=0.1$\,T (b) FeCo-122 samples. After ZFC, an external magnetic field was applied in $\approx 250$ G increments. (c) Comparison of two profiles obtained at $H=2$\,kOe. (d) Simultaneous magneto-optical imaging of two crystals at $H= 1$\,kOe.}
\label{fig2}
\end{center}
\end{figure}

Figure\,\ref{fig1} shows MO images and magnetic induction profiles obtained simultaneously after field cooling (FC) in a 2 kOe magnetic field and turning the field off at 5 K for FeCo-122 crystals. The amount of trapped flux, proportional to the persistent
(Bean) current in the sample, increases dramatically in the irradiated samples compared to the reference sample even at a
modest $B_{\phi}=0.1$\,T. At $B_{\phi}=2$\,T irradiation, the effect is even larger, but it is difficult to compare because
2 kOe is apparently less than a full Bean penetration field, $H^* \sim J_c$, at this temperature, so the profiles are
rounded at the sample center. In the reference sample, the full critical state is reached and we can fit $B_z(x)$
to a formula for a 2D superconducting film,
\begin{equation}
B_z(x)=\frac{J_{c}2d}{c}\ln{\left[\frac{(z^2+(x-a)^2)(z^2+(x+a)^2)}{(z^2+x^2)^2}\right]}
\label{eq1}
\end{equation}
\noindent where $z$ is the distance from the surface to the MO indicator. The fit with $J_c$(15 K)$\approx 0.108$ MA$\cdot$cm$^{-2}$ is shown by a solid line in Fig.\,\ref{fig1}. From the magnetization measurements we obtained $J_c(15 K, H=0)\approx 0.094$ MA$\cdot$cm$^{-2}$, so the agreement is quite good. The difference in the shielding ability is clearly seen in the data shown in Fig.\,\ref{fig2}. Profiles of $B(r)$, measured after cooling in zero field (ZFC) to 15 K and applying a magnetic field at roughly 250 Oe increments are shown for the reference (a) and irradiated, $B_{\phi}=0.1$\,T (b) samples. Panel (c) compares two profiles at 2 kOe where substantial enhancement of pinning is observed in the irradiated sample. The MO images (d) show partial flux penetration at 1 kOe demonstrating a much stronger, \emph{spatially uniform}, shielding in the irradiated sample.

\begin{figure}[tb]
\begin{center}
\includegraphics[width=8.4cm]%
{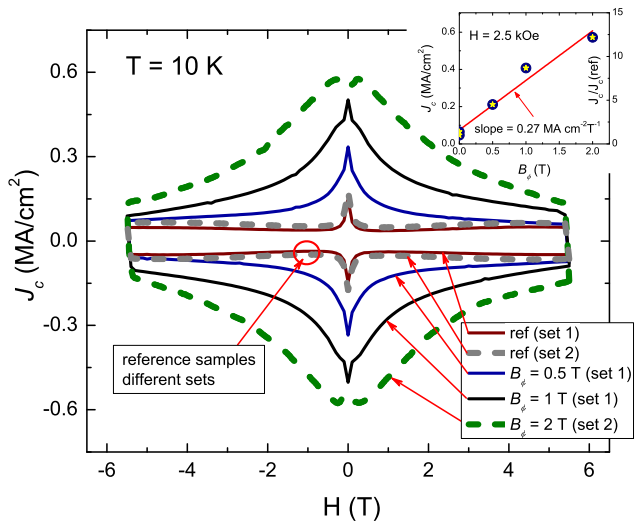}%
\caption{(Color online) Magnetization loops (converted into $J_c$ using Bean model) measured at 10 K in two sets of FeCo-122 single crystals with: set \#1 (solid lines) $B_{\phi}=0,\,0.5$\,T and $B_{\phi}=1$\,T and set \#2 (dotted lines) $B_{\phi}=0$\,T and $B_{\phi}=2$\,T. Inset: $J_c$ (left axis) and $J_c/J_c(ref)$ (right axis) at $H=2.5$ \,kOe vs. $B_{\phi}$. The solid line is a linear fit with a slope of 0.27 MA$\cdot$cm$^{-2}$T$^{-1}$.}
\label{fig3}
\end{center}
\end{figure}
\begin{figure}[tb]
\begin{center}
\includegraphics[width=8.5cm]{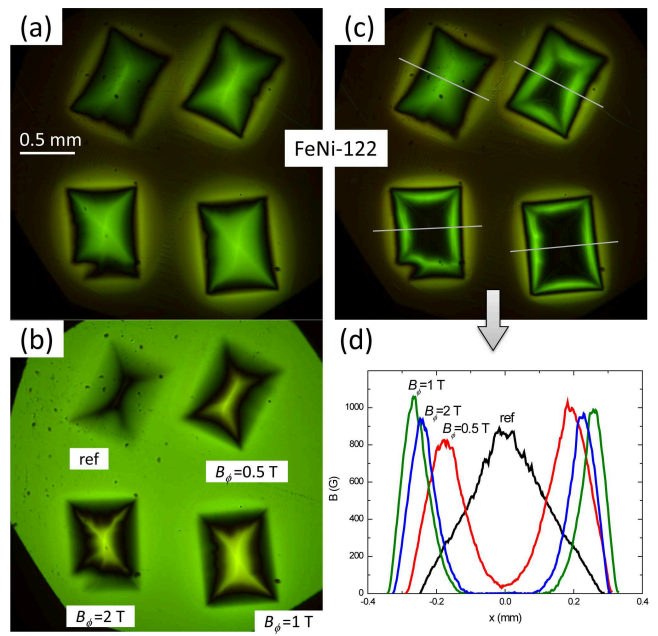}
\caption{(Color online) Irradiated FeNi-122 crystals with $B_{\phi}=0.5\,1$ and $2$\,T imaged at 5 K simultaneously with the reference sample. (a) Trapped flux after 2 kOe FC from above $T_c$ and turning it off at 5 K (similar to Fig.\ref{fig1} for FeCo-122). (b) Penetration of the antiflux shown for $H=-800$ Oe (of opposite sign to the trapped, shown in (a)). (c) Trapped flux after partial penetration after applying 2 kOe field and turning it off. (d) $B(r)$ along the directions shown in (c).}
\label{fig4}
\end{center}
\end{figure}

The MO measurements are in agreement with the measurements of macroscopic magnetization converted into the critical current density shown for FeCo-122 crystals in Fig.\,\ref{fig3} (a small \emph{reversible} and reproducible $M(H)$ background was measured at $T>T_c$ and subtracted). The figure shows data for two sets of FeCo-122 systems, altogether covering all available irradiation doses, $B_{\phi} = 0,~0.5,~1,~2$ T. Evidently, there is a large overall enhancement of the irreversible magnetic moment. The inset in Fig.\,\ref{fig3} shows the critical current density, $J_c$, measured at 2.5 kOe (to avoid effects discussed below, yet stay at $H < B_{\phi}$) plotted as a function of the $B_{\phi}$. Unirradiated samples from both sets show very similar $J_c$(15 K, 2.5 kOe) $\approx 0.05$ MA/cm$^2$, whereas irradiation enhances $J_c$ at a rate of about 0.27 MA$\cdot$cm$^{-2}$T$^{-1}$, or in other words, at $B_{\phi}=2$\,T, $J_c$(15 K,2.5 kOe) is about 12 times larger than in the unirradiated sample. At larger fields, the effect becomes weaker as expected for $H \gg B_{\phi}$ when the irradiation-produced defects saturate with vortices. Due to randomness of the damage, nothing happens exactly at $B_{\phi}$. Also note the disappearance of the fishtail (second magnetization peak) upon irradiation. In the literature, $J_c$ is often reported in the remanent state at 5 K. In pnictides, an anomalously large peak is observed in $M(H)$ loops at $H \rightarrow 0$ and it is not clear whether it has intrinsic or aspect ratio - related (or both) origin. Irradiation affects this peak and it is possible that the different shapes of $B(r)$ in Fig.\,\ref{fig1} are reflecting this effect. From our measurements at 5 K in $H=0$, the current density increases from $J_c$(5 K, 0) $\approx 0.30$ MA/cm$^2$ to about $J_c (5 K, 0) \approx 0.95$ MA/cm$^2$, or by only a factor of three. However, already at 2.5 kOe (beyond the peak location), it changes from from $J_c$(5 K, 2.5 kOe)$\approx 0.16$ MA/cm$^2$ to about $J_c$(5 K, 2.5 kOe)$\approx 0.92$ MA/cm$^2$, or a factor of almost six. Yet, this enhancement is almost two times lower than at 15 K, Fig.\,\ref{fig3}. For comparison, in the remanent state at 5 K, neutron irradiation reported a three-fold increase\cite{Zehetmayer2009a}, whereas 200 MeV Au ions irradiation resulted a six-fold enhancement.\cite{Nakajima2009}

Finally, Fig.\,\ref{fig4} shows a similar set of measurements on four \FeNi single crystals, again all parts of the same large, original crystal. Figure\,\ref{fig4}(a) presents MO imaging of the trapped flux obtained after 2 kOe FC to 5 K and turning the field off. The intensity is proportional to the amount of trapped flux and, while the irradiation effect is clear with respect to the unirradiated sample, the difference between the $B_{\phi}=1$\,T and $B_{\phi}=2$\,T samples is not obvious. It becomes apparent that pinning enhancement is \emph{larger} in the $B_{\phi}=1$\,T sample by examining Fig.\,\ref{fig4}(b) that shows penetration of the antiflux (magnetic field of opposite sign compared to the trapped flux). In this experiment switching of the field sign inside the sample outlines the boundary of $B=0$. While $H=-800$ Oe has almost fully penetrated the unirradiated sample, the penetration is only partial in the case of irradiated crystals. While pinning in the sample with $B_{\phi}=0.5$\,T seems comparable to that with $B_{\phi}=2$\,T, the sample with the $B_{\phi}=1$\,T shows the largest pinning. Another "high - contrast" measurement is shown in Fig.\,\ref{fig4}(c) and (d). After ZFC to 5 K, a magnetic field of 2 kOe was applied and removed. It fully penetrated the unirradiated sample, but only partially the irradiated crystals. The width of the penetrated "Bean belt" is proportional to $J_c$ and is easy to measure from the profiles. Even visually it is obvious that the sample with $B_{\phi}=2$\,T shows pinning larger than with $B_{\phi}=0.5$\,T, but smaller than the sample with with $B_{\phi}=1$\,T. It is more clearly seen in the profiles of $B(r)$ shown in Fig.\,\ref{fig4}(d) where the latter sample has the smallest width of penetration and largest maximum, both yielding a larger gradient, $dB/dr$, proportional to $J_c$.

The gross interpretation of our results is straightforward. Heavy - ion irradiation leads to a \emph{substantial}, \emph{spatially uniform} enhancement of flux pinning at least in the optimally-doped crystals of the FeT-122 family of iron-based pnictide superconductor (In the underdoped regime structural/magnetic domains may interfere with the additional damage produced by heavy ions). While we do not (yet) have electron-microscopy confirmation of the type of the defects produced by the ions, the induced defects are certainly effective pinning centers. A detailed investigation of the magnetic, thermodynamic and transport responses of the irradiated crystals will be published elsewhere. Here our results show that the pinning enhancement is large and mesoscopically uniform. However, it appears that larger irradiation doses, $B_{\phi}>1$\,T, are not so effective in FeNi-122, but are still effective in FeCo-122. While pinning in unirradiated samples is almost identical in these two systems, the higher-dose irradiation may start to suppress superconductivity in FeNi-122 due to its lower, overall weaker superconductivity, judged by its lower $T_c$ and larger London penetration depth.


\section{Acknowledgements}

We thank R. T. Gordon for help in improving the manuscript. Work at the Ames Laboratory was supported by the Department of Energy  - Office of Basic Energy Sciences under Contract No.\,DE-AC02-07CH11358. Work at Argonne National Laboratory was supported by the U.S. Department of Energy, Office of Science, Office of Basic Energy Sciences under contract No. DE-AC02-06CH11357. The heavy ion irradiation was performed at the ATLAS facility at Argonne. R.P. acknowledges support from the Alfred P. Sloan Foundation.


%

\end{document}